\documentclass{ws-procs9x6-cpt19}
\begin{document}

\def\etal {{\it et al.}}

\title{Developments in Lorentz and CPT Violation}

\author{V.\ Alan Kosteleck\'y}

\address{Physics Department, Indiana University\\
Bloomington, IN 47405, USA}

\begin{abstract}
This talk at the CPT'19 meeting outlines 
a few recent developments in Lorentz and CPT violation,
with particular attention to results obtained by researchers 
at the Indiana University Center for Spacetime Symmetries.
\end{abstract}

\bodymatter

\section{Introduction}

Motivated by the prospect of minuscule observable effects 
arising from Planck-scale physics,
searches for Lorentz and CPT violation 
have made impressive advances in recent years.\cite{tables}
The scope of ongoing efforts presented at the CPT'19 meeting
indicates that this rapid pace of development 
will continue unabated,
with experiments achieving sensitivities to Lorentz violation 
that are orders of magnitude beyond present capabilities
and providing unprecedented probes of the CPT theorem.
Substantial theoretical advances in the subject are also being made,
and the prospects are excellent for completing 
a comprehensive description of possible effects 
on all forces and particles 
and for achieving a broad understanding 
of the underlying mathematical structure
in the near future.
In this talk,
I summarize some basics of the subject
and outline a few recent results obtained 
at the Indiana University Center for Spacetime Symmetries (IUCSS).

\section{Basics}

No compelling experimental evidence for Lorentz or CPT violation
has been reported to date,
so any effects are expected to involve only tiny deviations
from the physics of General Relativity (GR) and the Standard Model (SM).
In studying the subject,
it is therefore desirable to work within 
a theoretical description of Lorentz and CPT violation
that is both model independent
and includes all possibilities consistent 
with the structure of GR and the SM.
The natural context for a description of this type
is effective field theory.\cite{eft}
The realistic and coordinate-independent effective field theory 
for Lorentz and CPT violation is known as 
the Standard-Model Extension (SME).\cite{ck,akgrav}
It can be obtained by incorporating 
all coordinate-independent and Lorentz-violating terms
in the action for GR coupled to the SM.
These terms also describe general realistic CPT violation,\cite{ck,owg}
and they are compatible with either spontaneous or explicit Lorentz violation
in an underlying unified theory such as strings.\cite{ksp}

The SME action incorporates Lorentz-violating operators
of any mass dimension $d$,
with the minimal SME defined to include the subset 
of operators of renormalizable dimension $d\leq 4$.
A given SME term is formed as the observer-scalar contraction
of a Lorentz-violating operator with a coefficient for Lorentz violation
that acts as a background coupling to control observable effects.
The propagation and interactions of each species are modified  
and can vary with momentum, spin, and flavor.
All minimal-SME terms\cite{ck,akgrav}
and many nonminimal terms\cite{nonmin-sm,kl19,nonmin-gr}
have been explicitly constructed.
The resulting experimental signals are expected
to be suppressed either directly
or through a mechanism such as countershading 
via naturally small couplings.\cite{akjt}
Impressive constraints on SME coefficients from many experiments
have been obtained.\cite{tables}
The generality of the SME framework insures these constraints
apply to any specific Lorentz-violating model
that is consistent with realistic effective field theory. 

The geometry of Lorentz violation is an interesting issue for exploration.
If the Lorentz violation is spontaneous,
then the geometry can remain Riemann or Riemann-Cartan\cite{akgrav} 
and the phenomenology incorporates Nambu-Goldstone modes.\cite{lvng}
However, if the Lorentz violation is explicit,
then the geometry cannot typically be Riemann
and is conjectured to be Finsler instead.\cite{akgrav} 
Support for this idea has grown in recent years,\cite{finsler,ek18}
but a complete demonstration is lacking at present.

\section{Developments from the IUCSS}

In the past three years,
developments from the IUCSS have primarily involved 
the quark, gauge, and gravity sectors.
In the quark sector,
direct constraints on minimal-SME coefficients 
can be extracted using neutral-meson oscillations,\cite{ak98}
and numerous experiments on $K$, $D$, $B_d$, and $B_s$ mixing
have achieved high sensitivities to CPT-odd effects
on the $u$, $d$, $s$, $c$, and $b$ quarks.\cite{mesonexpt}
Recent work reveals that nonminimal quark coefficients at $d=5$ 
provide numerous independent measures of CPT violation,\cite{ek19}
many of which are experimentally unconstrained to date.
The $t$ quark decays too rapidly to hadronize,
but $t$-$\overline{t}$ pair production and single-$t$ production 
are sensitive to $t$-sector coefficients and are the subject
of ongoing experimental analyses.\cite{top} 
High-energy studies of deep inelastic scattering and Drell-Yan processes 
also offer access to quark-sector coefficients,\cite{klv16} 
and corresponding data analyses are being pursued.
Another active line of reasoning 
adapts chiral perturbation theory
to relate quark coefficients to hadron coefficients,\cite{lvchpt}
yielding further tests of Lorentz and CPT symmetry.

In the gauge sector,
the long-standing challenge of constructing
all nonabelian Lorentz-violating operators at arbitrary $d$
has recently been solved.\cite{kl19} 
The methodology yields all matter-gauge couplings,
so the full Lorentz- and CPT-violating actions 
for quantum electrodynamics, quantum chromodynamics, 
and related theories are now available for exploration. 
Constraints on photon-sector coefficients continue to improve.\cite{photons}
Signals of Lorentz violation arising in clock-comparison experiments
at arbitrary $d$ have recently been studied,\cite{kv18} 
revealing complementary sensitivities from 
fountain clocks,
comagnetometers,
ion traps,
lattice clocks,
entangled states,
and antimatter.
These various advances suggest excellent prospects for future searches 
for Lorentz and CPT violation in the gauge and matter sectors.

In the gravity sector,
all operators modifying the propagation 
of the metric perturbation $h_{\mu\nu}$,
including ones preserving or violating Lorentz and gauge invariance,
have been classified and constructed.\cite{nonmin-gr}
Many of the corresponding coefficients are unexplored but could be measured
via gravitation-wave and astrophysical observations.
A general methodology exists for analyzing Lorentz-violation searches
in experiments on short-range gravity,\cite{nonmin-gr}
and constraints on certain coefficients with $d$ up to eight 
have now been obtained.\cite{shortrange}
Work in progress further extends gravity-sector tests
to matter-gravity couplings at arbitrary $d$.\cite{kl19ip}
Results from the SME can also be applied 
to constrain hypothesized Lorentz-invariant effects
whenever these lead to nonzero background values for vector or tensor objects.
This idea 
recently yielded the first experimental constraints on all components
of nonmetricity.\cite{nonmetricity}
At the foundational level,
further confirmation of the correspondence 
between the SME and Finsler geometry has been established 
via the construction of all Finsler geometries 
for spin-independent Lorentz-violating effects.\cite{ek18}
The scope and breadth of all these results 
augurs well for future advances in the gravity sector
on both theoretical and experimental fronts.

\section*{Acknowledgments}

This work was supported in part
by U.S.\ D.o.E.\ grant {DE}-SC0010120
and by the Indiana University Center for Spacetime Symmetries.

\end{document}